# Bright Spot Characterization of Low dI/dt X-pinch Plasmas using Soft X-ray Spectroscopy with Bennett Relation

YeongHwan Choi,[1] Muhyeop Cha,[1] Hakmin Lee,[1] Hsiao-Chien Chi,[1] Seongmin Choi,[2] Seungmin Bong,[2] Seonghun Jeon,[1] Ookjin Choi,[1] Young-chul Ghim,[2] Yong-Seok Hwang,[1] and Kyoung-Jae Chung[1,a]

[1]*Department of Energy Systems Engineering, Seoul National University, Seoul, Republic of Korea*
[2]*Department of Nuclear and Quantum Engineering, KAIST, Daejeon, Republic of Korea*
[a]Correspondence: jkjlsh1@snu.ac.kr

This study investigates the characteristics of X-pinch plasmas driven under low current rise rate (dI/dt) conditions using soft x-ray spectroscopy combined with the Bennett relation. X-pinch experiments were conducted on the SNU X-pinch device using copper wires at a low dI/dt of 0.2–0.3 kA/ns. The resulting 1–10 keV soft x-ray signals, measured by an x-ray filtered AXUV photodiode array (XFPA), exhibit significant nonlinear effects due to the high intensity of the soft x-ray pulses. This work characterizes the nonlinear behavior of the AXUV-HS5 Si PIN photodiodes under intense pulsed radiation using a pulsed laser. We identified a charge conservation property that the total collected charge remains proportional to the incident pulse energy despite temporal profile distortion. Based on this diagnostic finding, we developed and applied a new framework for plasma parameter estimation. By combining a spherical emission model with the Bennett equilibrium, this approach determines that the soft x-ray source plasma is a 'bright spot', characterized by a plasma density $n_e \sim 10^{21}$ cm$^{-3}$, size d ~30–40 μm, electron temperature $T_e$ ~1 keV, and an emission duration $t_B$ ~1 ns, rather than an extremely compressed 'hot spot'.

## I. Introduction

An X-pinch, formed by two or more thin wires arranged to cross at a single point, is a compact source of intense soft x-ray bursts.[1] When driven by a high-current pulse, the large self-magnetic pressure near the crossing point can lead to the formation of a hot, dense and compact plasmas that emit in the 1–10 keV band on sub-nanosecond time scales. Because of its small source size and high brightness, the X-pinch has been widely used for high-resolution point-projection radiography and for probing fast plasma dynamics in high-energy-density physics (HEDP).[2,3] Since the emitted x-ray spectrum contains essential information about the extreme state of the source, diagnostics that can provide reliable, quantitative spectral data are necessary for probing X-pinch dynamics.

The dynamics of X-pinch plasmas are well established[1,4] on fast-rise drivers satisfying the Shelkovenko condition[1,5] (dI/dt > 1 kA/ns), which are generally required to generate compact, high-density (~$10^{23}$ cm$^{-3}$) 'hot spot' plasmas[6]. However, the behavior in the slow-rise regime (dI/dt < 1 kA/ns), often accessible with compact capacitor discharge platforms, has been less comprehensively characterized. Nonetheless, studies consistently show that even slow-rise X-pinches can generate useful, compact soft x-ray (SXR) sources.[7] Previous works indicate the formation of micro z-pinches emitting ~1–2 keV SXR bursts on ~1 ns timescales, even at current rise rates (dI/dt) well below 1 kA/ns[8-11]. Electron temperatures are typically inferred to be in the range of several hundred eV to ~1.5 keV[8,9]. Imaging results suggest source sizes typically in the range of tens of microns, with the apparent size depending on the x-ray filter's cutoff energy[9,12,13]. For instance, Strucka et al.[7], using a portable driver of ~0.45 kA/ns, measured reliable 1.1±0.3 ns bursts with soft-band spot sizes of ~10±6 μm, demonstrating the diagnostic potential of these low dI/dt X-pinch sources.

Quantitative interpretation in many of these studies relies on diagnostics like X-ray filtered photodiode arrays (XFPA), which offer spectral responsivity.[14-16] However, in practice, XFPAs in pulsed power experiments often record waveforms with extended tails that can persist for longer than 10 ns.[13,17] This duration cannot be accounted for by instrument bandwidth alone and is consistent with saturation-induced space charge (plasma) effects in the intrinsic region of a silicon PIN photodiode. This artifact undermines confidence in both the temporal response and the interpretation of X-pinch dynamics. While the potential for photodiode saturation has been noted, it has rarely been treated quantitatively as a primary measurement bias in X-pinch experiments.

In this study, we present a measurement and analysis methodology for extracting reliable plasma parameters from saturated XFPA signals produced by single thermal x-ray bursts in low dI/dt X-pinches. Using intensity controlled 150 ps pulsed laser tests on AXUV-HS Si PIN photodiodes, we characterize the onset of space-charge saturation, identify the origin of long signal tails, and determine the linear regime impulse response of the diagnostic system. Building on the observed charge conservation property, we then develop a charge-based inversion that couples the integrated current from each XFPA channel to a luminosity and opacity consistent spherical radiation model, and closes the inference with an independent Bennett relation constraint to obtain $n_e$, $T_e$, d, and the thermal burst emission duration $t_B$. Section II describes the device and diagnostics, and Section III details the laser-based characterization of the photodiode. Section IV formulates the charge-based inversion with radiation model and the Bennett relation. Section V applies the framework to copper X-pinch shots at $dI/dt$ = 0.2–0.3 kA/ns, and Section VI discusses assumptions (single burst), limitations, and implications for low $dI/dt$ drivers.

## II. Experimental Setup

Experiments with Cu-wire X-pinches were

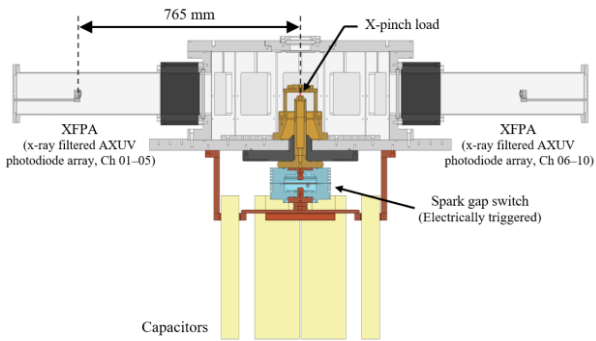

FIG. 1. Schematic of the SNU X-pinch with the x-ray filtered AXUV photodiode array (XFPA). The capacitor discharge is controlled by an electrically triggered spark gap switch. The XFPA comprises 10 channels, each with an AXUV-HS5 photodiode and a different x-ray filter. The distance between the X-pinch load and the XFPA is 765 mm.

performed on the SNU X-pinch device.[18] Figure 1 provides a schematic of the SNU X-pinch and its x-ray diagnostic system (XFPA). After several modifications, the SNU X-pinch device has an inductance of 140 nH and a resistance of 40 mΩ. The capacitor bank comprises eight capacitors, each with a capacitance of 0.1 μF, for a total of 0.8 μF. The maximum operating voltage is 80 kV, and the short-circuit current rise rate at a 50 kV charge voltage is ~0.25 kA/ns. A Cu-wire X-pinch experiment performed on the SNU X-pinch showed multiburst of x-rays, which were detected by the XFPA. Measured current and x-ray signals for shot 378 (50 kV, 25 μm wire) are presented in Figure 2.

The XFPA (X-ray Filtered AXUV Photodiode Array) was developed for the SNU X-pinch for quantitative soft x-ray spectroscopy, covering the energy range of 1–10 keV[19]. The distance between the XFPA and X-pinch load is 765 mm, and all ten channels face the wire-crossing point. Each channel includes a foil filter with a spectral edge in the 1–10 keV range, and a reverse biased AXUV-HS5 photodiode with a bias voltage of 50 V. The spectral response of the AXUV-HS5 silicon photodiode is presented in Figure 3(a), showing the calculated response based on a 100 μm effective silicon thickness.[19] For the evaluation of deposited energy, we used mass absorption coefficients from Gullikson et al.[20] for 0.4–10 keV and from NIST[21] for 1–30 keV. The calculated response agrees well with previous Geant4 simulations[19] and other studies on AXUV-HS5 photodiodes[1, 22]. The AXUV-HS5 responsivity is constant at 0.27 A/W for x-rays below 4 keV and decreases to nearly zero above 20 keV. The spectral response curves of all XFPA channels, shown in Figure 3(b), were obtained by multiplying the photodiode response with the transmittance of the corresponding x-ray filters. The XFPA does not respond to x-rays below 0.8 keV, and all channels except Ch 02 respond only to x-rays above 1 keV. Details of the filter configurations are available in Ham et al.[19], and filter transmittances were calculated using data from Henke et al.[23]

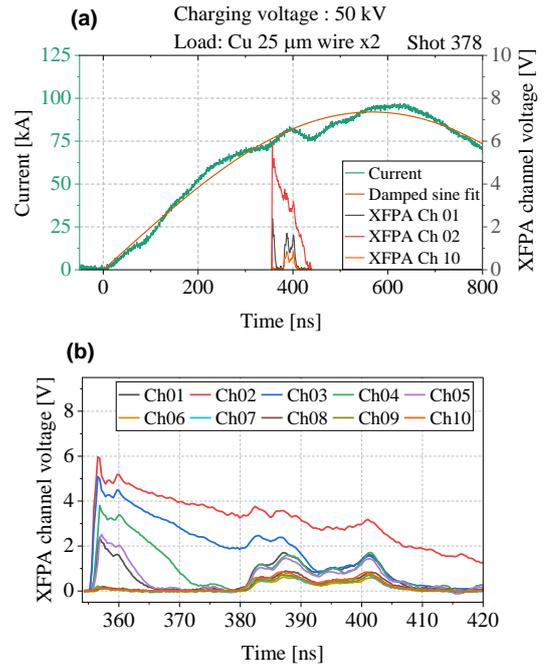

FIG. 2. Current and x-ray signals from a representative X-pinch experiment (Shot 378). A charging voltage of 50 kV and a 25 μm diameter two-wire copper load were used. The start of the current rise is set to $t = 0$. (a) Load current and representative XFPA signals from Ch 01, 02, and 10. (b) Temporal responses of all ten XFPA channels.

The XFPA is designed to selectively detect radiation from the high-energy-density plasma core. Early-stage corona plasmas, with densities of $10^{17}$–$10^{19}$ cm$^{-3}$ and temperatures of 10–100 eV, do not emit significantly in the SXR range[1]. According to Aranchuk et al.[13], photons above 300 eV appear only at the main-burst stage. This implies that the XFPA, which is sensitive only to x-rays above 0.8 keV, effectively isolates the emission from the dense, hot stages of the X-pinch evolution.

## III. Characterization of Photodiode Nonlinearity

Initial analysis of XFPA data from the SNU X-pinch suggested that the output of low-energy channels, specifically Ch 02 (>0.8 keV) and Ch 03 (>1 keV), was not proportional to the incident x-ray flux under high-intensity conditions. Figure 4 shows the XFPA signals from a copper wire X-pinch experiment (Shot 549). At the signal peak near 331 ns, both Ch 02 and Ch 03 exhibit a nearly identical peak voltage of ~8.5 V. This result is unphysical given their distinct spectral response curves, indicating a nonlinear detector response above a certain flux threshold. A rough estimate of the incident x-ray power for this shot suggests a peak flux that would generate a photocurrent corresponding to tens of volts in the linear regime. This is far above the ~1 V saturation threshold determined from our laser calibration experiments, providing strong evidence that the detector was operating deep within the saturated regime.

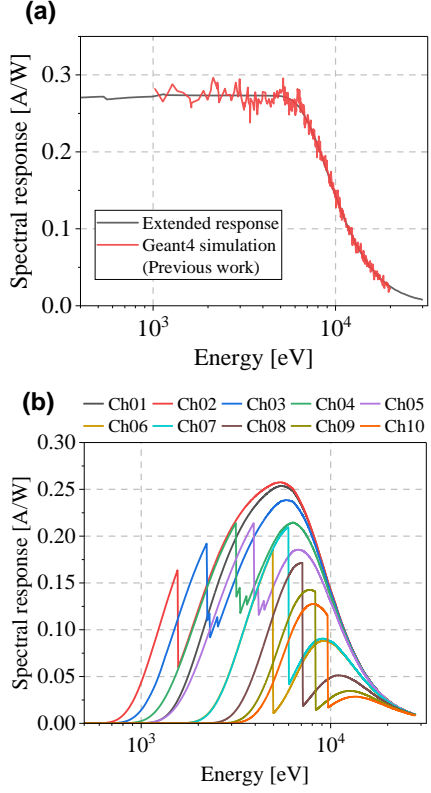

FIG. 3. Spectral responses of the AXUV-HS5 photodiode and the XFPA channels. (a) The calculated spectral response of a bare AXUV-HS5 photodiode, which corresponds well with previous Geant4 simulations. (b) The spectral responses of the 10 XFPA channels, where the response of the photodiode is multiplied by the transmittance of each x-ray filter.

To investigate this nonlinearity, we performed an experiment to characterize the photodiode's temporal response and saturation properties using a 532 nm, 150 ps pulsed laser (EKSPLA, SL234-SH-10). The experimental setup is shown in Figure 5(a). The laser pulse energy was precisely controlled using a stack of absorptive neutral-density (ND) filters. The AXUV-HS5 photodiode was configured identically to the X-pinch experiments, with a -50 V reverse bias, and connected to a 1 GHz oscilloscope with a 50 Ω input impedance. Since the 150 ps laser pulse is much shorter than the ~700 ps rise time of the photodiode, it acts as an impulse input, allowing the measured signal in the linear regime to be treated as the system's impulse response.

Figure 5(b) shows the measured voltage waveforms at different attenuation levels. For attenuations less than -45 dB, a long tail extending beyond 10 ns appears, growing to over 30 ns at -40 dB. The peak voltage, plotted in Figure 5(c), follows the expected linear scaling for attenuations stronger than -55 dB. However, for more intense pulses, the peak voltage deviates from this scaling, indicating the onset of nonlinearity. A peak voltage of ~1 V can be considered the threshold for the AXUV-HS5 photodiode in our 50 Ω transimpedance system.

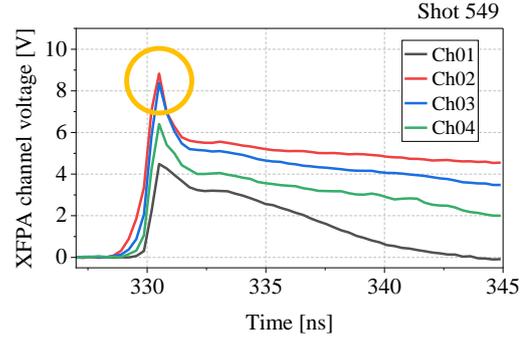

FIG. 4. XFPA voltage signals from a copper wire X-pinch (Shot 549) showing evidence of saturation. The nearly identical peak voltages of Ch 02 and Ch 03 at ~8.5 V are unphysical given their distinct spectral responses, indicating a nonlinear detector response.

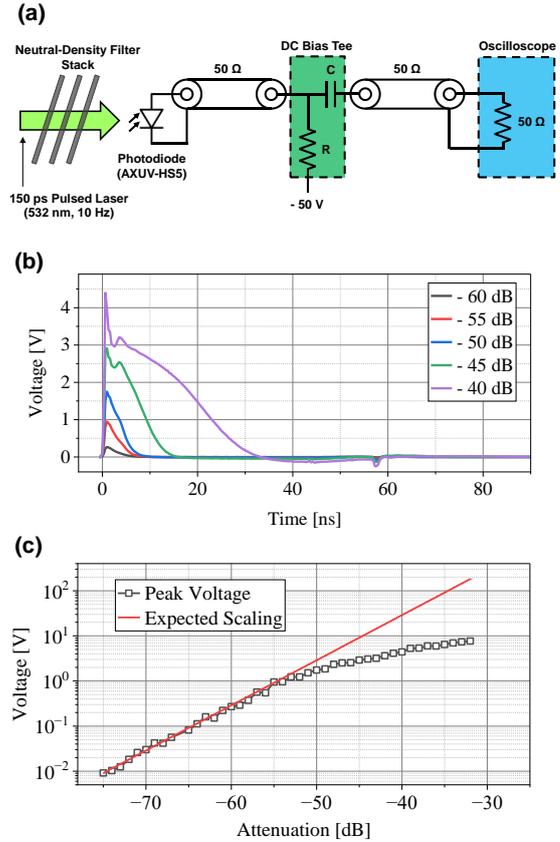

FIG. 5. Characterization of photodiode saturation using a 150 ps pulsed laser. (a) Schematic of the experimental setup for measuring the photodiode response to controlled-intensity laser pulses. (b) Measured voltage waveforms for different laser pulse energies. (c) Peak output voltage as a function of pulse energy. The deviation from the expected linear scaling (red line) starts at ~1 V.

This nonlinear behavior in semiconductor detectors under high, pulsed flux is primarily caused by space-charge effects (also called plasma effects)[24, 25]. A high density of photo-generated carriers screens the internal electric field, which slows the carrier drift velocity and prolongs the charge collection time. Alternative causes for the long signal tail, such as impedance mismatch or load line effect, were also considered. However, the observed strong dependence of the tail duration on the incident pulse intensity is a classic signature of plasma effects and space-charge dynamics, making it the most plausible explanation.[26, 27] While previous works[17, 28] have reported on nonlinear operation (e.g. formation of a tail) in Si PIN photodiodes in pulsed-power environments, these studies have mainly focused on modeling or provided qualitative discussions, without quantitatively addressing saturation behavior.

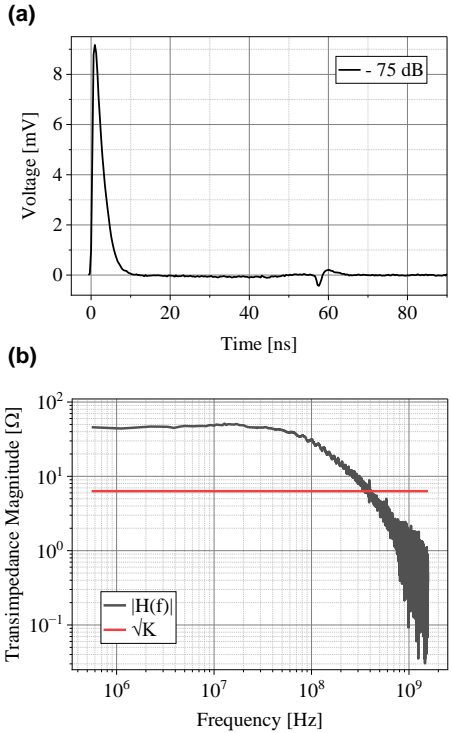

FIG. 6. Linear response of the diagnostic system. (a) Impulse response measured with a weak (-75 dB) laser pulse where space-charge effects are negligible. The 4.5 ns fall time is consistent with the hole transit time across the 100 µm intrinsic region. (b) The system transfer function, |H(f)|, derived from the impulse response, showing the transimpedance as a function of frequency. The square root value of the Tikhonov–Wiener regularization constant K is also shown.

The carrier generation profile produced by 532 nm light, which is absorbed within the first few microns of the silicon surface[29], differs from that of 1–10 keV x-rays, which deposit energy more uniformly throughout the 100 µm intrinsic region. However, despite these differences in carrier generation dynamics, previous studies[15, 30] have shown that the time response and total collected charge (proportional to the incident pulse energy) of silicon PIN photodiode under an x-ray pulse closely matches those obtained with pulsed laser experiments. Therefore, this laser-based experiment provides clear evidence of the saturation phenomenon and enables the characterization of the linear regime impulse response, which is dominated by carrier transit times and the readout bandwidth. The impulse response of the complete diagnostic system, measured in the linear regime (-75 dB), is shown in Figure 6(a). The waveform exhibits a 0.7 ns rise time and a 4.5 ns fall time. The slow decay is consistent with the calculated hole transit time across the 100 µm intrinsic region (t ~ $d^2/\mu_p V$ ~ 4.4 ns), confirming that space-charge effects are negligible in this regime. The transfer function, H(f), derived from this impulse response is shown in Figure 6(b).

Using this transfer function, we can reconstruct the effective source current, I(f), from the measured voltage, V(f), using Wiener deconvolution:

$$I(f) = V(f) \frac{H^*(f)}{|H(f)|^2 + K} \qquad (1)$$

where $H^*(f)$ is the complex conjugate of H(f) and K is a Tikhonov–Wiener regularization constant with units of $\Omega^2$ (here, K = 40 $\Omega^2$). We determined K by sweeping from smaller to larger values and selecting the smallest K that suppresses unphysical Gibbs-like ringing around steep edges and preserves the integrated charge. We also confirmed that varying K within a reasonable range (20-80 $\Omega^2$) does not change the reconstructed effective source current profile significantly. The effective source current represents the equivalent input that would produce the measured output voltage if the system response remained linear. For non-saturated signals (e.g., -60 dB), the reconstructed waveform accurately represents the incident radiation profile, as shown in Figure 7(a). In the linear regime (attenuation > -60 dB), the method accurately reconstructs the ~1 ns FWHM temporal profile of the incident laser pulse. However, for saturated signals (Figure 7(b)), it reveals a compressed peak and a long tail, quantifying the temporal distortion introduced by the saturation dynamics as an equivalent input.

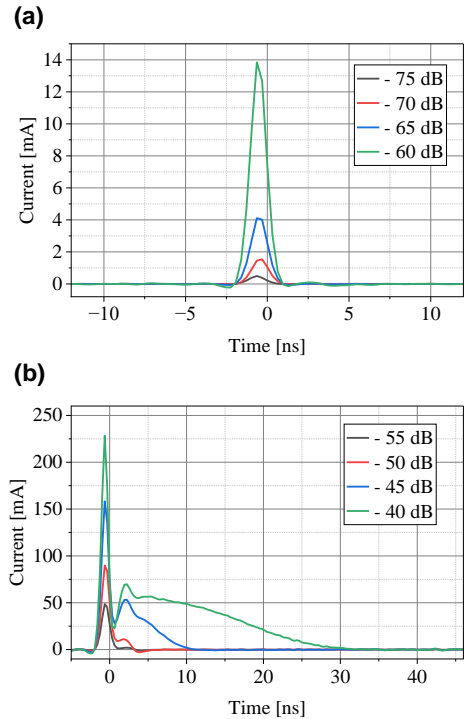

FIG. 7. Reconstructed effective source current using Wiener deconvolution. (a) In the linear regime. (b) In the saturated regime, the reconstruction reveals a compressed peak and a long tail.

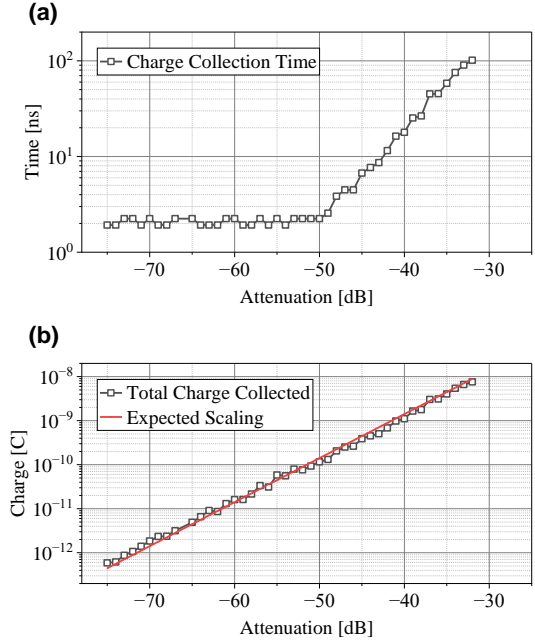

FIG. 8. Quantitative analysis of saturation effects on charge collection. (a) The 90% charge collection time, derived from the temporal profile of effective source current. (b) The total collected charge (the time integral of the effective source current) remains linear with respect to the incident pulse energy.

The impact of saturation on the charge collection dynamics is summarized in Figure 8. The charge collection time increases dramatically from ~2 ns in the linear regime to ~100 ns for a -32 dB pulse (Figure 8(a)). Even with this severe temporal distortion, the collected charge (the time integral of the effective source current) remains linear with the incident pulse energy across the entire tested range (Figure 8(b)). This indicates that charge loss due to recombination is negligible, even under heavy saturation (peak voltage of ~10 V). Therefore, for single-pulse events, the incident energy can be reliably quantified from the total collected charge, even when the temporal information is lost.

## IV. Analysis Methodology

As established in Sec. III, while the temporal information from AXUV-HS5 photodiodes can be distorted when saturated by intense x-ray pulses, the total collected charge is conserved and remains proportional to the incident x-ray energy. With unreliable time resolution, the finding that the total collected charge is conserved provides a pathway to extract physical information from saturated signals. Therefore, this section presents a new analysis methodology to estimate the plasma parameters of the x-ray source based on the total collected charge in each XFPA channel.

This methodology comprises three components. First, the x-ray emitting source is described by a uniform spherical plasma model that consistently accounts for luminosity and opacity effects (Sec. IV.A). Second, this model is compared with the measured total effective source charge per channel to find the space of possible plasma parameters ($n_e$, $T_e$, d) (Sec. IV.B). Finally, the Bennett equilibrium relation is applied as a physical constraint to determine all parameters, including the burst duration ($t_B$) (Sec. IV.C). For clarity, this analysis focuses on single, thermal SXR burst events where non-thermal emission is negligible.

### A. Opacity and Luminosity Consistent Spherical Plasma Radiation Model

The high-energy-density core of an X-pinch that emits x-rays is optically thick[31] to visible and UV radiation but is translucent to higher-energy photons[32], enabling SXR spectroscopy of the dense core. However, the x-ray emitting region can be marginally optically thick ($\tau \sim 1$), especially for L-shell lines.[33-35] Reabsorption (opacity) can affect the observed 1–10 keV spectrum and should be included in the spectroscopic model[36]. A previous study by Ham et al.[37] accounted for opacity by assuming fixed plasma sizes (10 μm for a hot spot, 1 mm for electron-beam sources) and a slab geometry. This approach is limited because the plasma size is not known a priori, and slab geometry is less appropriate[38] for an x-ray source formed by rapid radial compression[32].

Here, we model the x-ray emitting source as a uniform, spherical plasma, as shown in the schematic in Figure 11(a). This assumption is justified on several grounds. First, time-resolved x-ray pinhole images of low-current X-pinches often show compact, quasi-spherical emitting regions[39, 40]. Second, the radial nature of the compression process and the common assumption of isotropic emission[7] are consistent with a spherical approximation. Finally, a spherical model provides a simple analytical framework that consistently links the source's size-dependent opacity and its absolute luminosity. While we acknowledge that the actual plasma likely has spatial gradients, this model allows for the estimation of spatially-averaged, effective parameters that characterize the emitting volume.

For a uniform spherical plasma of diameter d, the spectral flux $F_\nu$ at a distance L has an analytical solution[41] given by

$$F_\nu = \pi \left(\frac{\epsilon_\nu}{\kappa_\nu}\right)\left(\frac{d}{2L}\right)^2 \left(1 + \frac{2(\tau_0 + 1)e^{-\tau_0} - 2}{\tau_0^2}\right) \quad (2)$$

where $\epsilon_\nu$ is the volume emissivity, $\kappa_\nu$ is the opacity, and L is the distance to the detector (765 mm in this setup). $\tau_0$ is a maximum optical depth, which is $\tau_0 = \kappa_\nu d$. Equation (2) incorporates the effects of increasing optical depth and luminosity as the plasma volume expands. This approach enables parameter estimation that complements analyses based solely on spectral ratios[37], which are inherently insensitive to absolute luminosity.

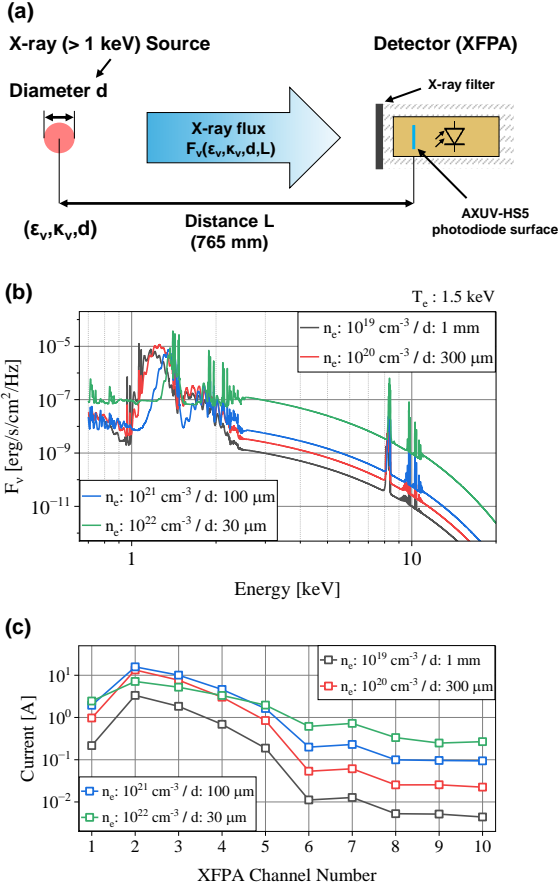

FIG. 9. (a) The schematics of opacity and luminosity consistent spherical plasma model. (b) Calculated x-ray spectral flux for a 1.5 keV plasma, showing the transition from L-shell line dominance to continuum dominance as density increases and size decreases. (c) The corresponding 10-channel XFPA currents for the spectra in (b), demonstrating the model's sensitivity to plasma parameters.

In this analysis, the volume emissivity $\varepsilon_v$ and opacity $\kappa_v$ are determined by three plasma parameters: electron density ($n_e$), electron temperature ($T_e$), and diameter (d). We calculated $\varepsilon_v$ and $\kappa_v$ for various combinations of these parameters using the FLYCHK[42] and FLYSPECTRA codes, which are widely used for analyzing radiation properties in high density x-ray emitting plasmas[36, 43]. We assume a pure copper (Z = 29) plasma. The calculation of $\varepsilon_v$ and $\kappa_v$ in FLYCHK requires a plasma size to account for self-absorption, where photon reabsorption alters the plasma composition (i.e., energy level and ion populations). Although FLYCHK can model this effect, it is limited to slab geometry. To maintain a one-zone spherical model while including self-absorption, we set the optical path length in FLYCHK to the mean escape length of a photon isotropically emitted from a uniform sphere, which is (3/8)d. This effective length approximation is employed because FLYCHK, as a 0D code, does not solve for non-local radiative transport. Therefore, the parameters derived using this method should be interpreted as effective, volume-averaged values.

The parameter space for the FLYCHK calculations was defined as follows: $n_e$ ranged from $10^{17}$ to $10^{25}$ cm$^{-3}$ and d from 0.1 µm to 1 cm, with 60 logarithmically spaced points selected for each. For the electron temperature $T_e$, 60 points were chosen between 0.1 and 3.0 keV at 48.3 eV intervals. This resulted in a total of $60^3$ = 216,000 parameter sets of ($n_e$, $T_e$, d) for which $\varepsilon_v$ and $\kappa_v$ were computed for x-ray energies from 0.7 to 28 keV. Since this analysis focuses on single thermal bursts, the fast electron fraction considered in Ham et al.[37] was excluded. Figure 11(b) shows the calculated spectral flux $F_v$ for a plasma at $T_e$ = 1.5 keV, for four different sets of density and size that produce a similar total x-ray power. At lower densities (e.g., $10^{19}$ cm$^{-3}$, d=1 mm), the emission is dominated by L-shell radiation (1–1.5 keV). As the density increases and size decreases, the L-shell radiation shifts to higher energies and the continuum radiation is significantly enhanced.

This spectral variation translates into a distinct signature in the calculated XFPA channel currents, as shown in Figure 11(c). The calculated spectral flux is then convolved with the XFPA response to obtain the expected current for each channel:

$$I_{ch} = A_{ch} \int_{0.7 \text{ keV}}^{28 \text{ keV}} F_v(E) R_{ch}(E) dE, \quad (3)$$

where $A_{ch}$ is the photodiode active area (1 mm²) and $R_{ch}(E)$ is the spectral response of the channel (Figure 3(b)). Figure 11(c) shows the resulting channel currents for the spectra in Figure 11(b). For the low-density, large-size case ($10^{19}$ cm$^{-3}$, 1 mm), the current is concentrated in the low-energy channels (Ch 02–04). Conversely, for the high-density, compact case ($10^{22}$ cm$^{-3}$, 30 µm), the currents in the higher-energy channels (Ch 01,05–10) significantly increase due to the enhanced continuum. The fact that each set of plasma parameters ($n_e$, $T_e$, d) produces a measurably different 10-channel current distribution forms the basis of our analysis. This sensitivity allows us to determine the effective plasma parameters by finding the set of model parameters that best fit the experimentally measured channel currents.

## B. Estimation of Plasma Parameters with Radiation Model

We estimate the plasma parameters of a single thermal burst by finding the set of ($n_e$, $T_e$, d) from our spherical radiation model that best reproduces the experimentally measured total charge collected in each of ten XFPA channels. Since the model calculates instantaneous channel currents ($I_{ch,model}$) based on ($n_e$, $T_e$, d), while the experiment provides total collected charge ($Q_{ch,exp}$), we need to assume the unknown burst duration $t_B$. Assuming a constant emission over this short duration, the modeled charge is $Q_{ch,model}(n_e,T_e,d,t_B)$ = $I_{ch,model}(n_e,T_e,d) \times t_B$. As established in Sec. III, the total collected charge $Q_{ch,exp}$, which is the time integral of the effective source current, is conserved even under heavy saturation. We therefore define an effective average channel current from the experiment as $I_{ch,exp}=Q_{ch,exp}/t_B$. The goal is then to find the ($n_e$, $T_e$, d) that minimizes the difference between $I_{ch,model}$ and $I_{ch,exp}$ for a given assumed $t_B$.

A least-squares (LS) method is employed to minimize the sum of squared differences between the ten model-calculated currents ($I_{ch,model}$) and the experimentally derived average currents ($I_{ch,exp}$). A direct LS fitting on these current values would be heavily biased by channels with large signals (e.g., Ch 02), while channels with signals near the noise floor (e.g., Ch 06–10 for typical thermal bursts) would have negligible influence. To ensure all channels contribute appropriately to the fit, the minimization is performed on the logarithm of the current values. To handle potential issues when taking the logarithm of values near zero or negative due to noise (especially after background subtraction), we employ a log-transformed least-squares fitting incorporating a stabilization constant ε. Specifically, any negative measured charges, which are actually below the noise floor, are treated as zero initially. Then, the positive constant ε, set to the experimentally determined noise floor (illustrated in Figure 11(a)), is added to both $I_{ch,model}$ and $I_{ch,exp}$ before taking the logarithm for the LS minimization. This approach effectively limits the influence of noise-dominated channels while preserving the spectral information contained across all ten channels. The robustness of this method was confirmed by verifying that varying ε over a reasonable range (e.g., 0.1 to 10 times the noise floor current equivalent) did not significantly alter the resulting best-fit parameters.

This parameter estimation, inherently comparing absolute current levels, directly accounts for the source's total luminosity. To characterize the uncertainty and potential degeneracies in the solution, we identify not only the single best-fit point but also a cluster of suboptimal solutions within a certain tolerance such that the sum of squared differences of the logarithms is within 3% of the minimum value. The center of mass of this cluster in the 3D parameter space ($n_e$, $T_e$, d) provides a practical estimate for the effective parameters, while the spread represents the solution confidence region. The shape and orientation of this region, which will be visualized later (see Fig. 14 in Sec. V), reveal the sensitivity of the fit and correlations between parameters.

This fitting procedure yields an optimal set of ($n_e$, $T_e$, d) for each assumed value of the burst duration $t_B$. To explore this dependency, we perform the fitting across a wide range of plausible soft x-ray burst durations of 30 ps to 3 ns. The results (presented in detail in Sec. V, e.g., Figure 11(b) for Shot 551) show that the estimated plasma density $n_e$ and electron temperature $T_e$ are relatively insensitive to the choice of $t_B$. This stability indicates that these parameters are constrained by the spectral shape, primarily determined by the ratios of the channel charges. However, the estimated source diameter d exhibits a clear dependence on the assumed $t_B$. This arises since, with $n_e$ and $T_e$ largely constrained by spectral ratios, the diameter d tends to vary to match the absolute luminosity required by the total measured charge and assumed duration $t_B$.

### C. Determination of Soft X-ray Burst Duration with Bennett Relation

As shown in Sec. IV.B, while the spectral shape derived from the ratios of the total collected charge across the channels constrains the $n_e$ and $T_e$, the absolute source size d and the underlying burst duration $t_B$ remain coupled through the total luminosity. To determine $t_B$, an independent physical constraint related to the plasma state during emission is required. For Z-pinch plasmas, including the micro Z-pinch formed at the neck of an X-pinch during peak compression, the Bennett relation provides such a constraint by describing the condition for quasi-equilibrium between the magnetic pressure generated by the pinch current and the thermal pressure of the plasma column.[44]

The applicability of the Bennett equilibrium model to the SXR emitting region in X-pinches, particularly around the time of peak compression and emission, is well-supported by previous X-pinch studies. Both theoretical models and simulations indicate that the hot spot formation process occurs under conditions approaching Bennett equilibrium[45-47]. Furthermore, experimental analyses across various current regimes, including slow-rise drivers, have successfully utilized Bennett-like relations to interpret observed scaling laws and plasma parameters[1, 11, 12, 48]. Although the X-pinch neck is a highly dynamic system, the SXR burst originates from the phase of local compression called cascading[1], during which a transient pressure equilibrium can be reasonably assumed.[48]

The Bennett relation which connects the pinch current ($I_p$) required to confine a plasma column with its line density and temperature is given as follows

$$2N_i k_B (Z^* T_e + T_i) = \frac{\mu_0 I_p^2}{4\pi} \quad (4)$$

where $N_i$ is ion line density, and $Z^*$ is the average ion charge. FLYCHK calculations show that for the plasma parameter range obtained by assuming the burst duration ($\log_{10}(n_e)$ ~ 21.4, $T_e \geq 1$ keV), the average ion charge $Z^*$ is conservatively estimated to be greater than 20. Therefore, assuming an isothermal plasma ($T_e \sim T_i$), the contribution from ion pressure becomes negligible compared to electron pressure ($Z^* T_e \gg T_i$), and the Bennett relation can be simplified in terms of the electron line density $N_e = Z^* N_i$.

$$2 N_e k_B T_e \sim \frac{\mu_0 I_p^2}{4\pi} \quad (5)$$

Here, $N_e$ is related to the plasma density $n_e$ and pinch diameter d by $N_e = n_e \times \pi(d/2)^2$.

In our analysis framework, for each assumed burst duration $t_B$, the fitting procedure described in Sec. IV.B yields a corresponding set of effective plasma parameters ($n_e$, $T_e$, d). From this set, we can calculate the electron line density $N_e(t_B) = n_e(t_B) \times \pi[d(t_B)/2]^2$. Using these values, we can determine the theoretical 'Bennett current' $I_{Bennett}(t_B)$ that would be required to confine such a plasma:

$$I_{Bennett}(t_B) = \sqrt{\frac{8\pi k_B T_e(t_B) N_e(t_B)}{\mu_0}} \quad (6)$$

The physically consistent solution occurs when the current required for equilibrium matches the actual current flowing through the pinch ($I_{p,measured}$) at the moment of the SXR burst. We measure $I_{p,measured}$ experimentally using Rogowski coils. By calculating $I_{Bennett}(t_B)$ over the range of assumed burst durations, we can identify the specific duration $t_B^*$ where $I_{Bennett}(t_B^*) = I_{p,measured}$. This single intersection point, illustrated for a representative shot in Sec. V (Fig. 14), enables the determination of the soft x-ray burst duration, thereby constraining all four effective parameters ($n_e$, $T_e$, d, $t_B^*$) within the model assumptions.

## V. Results and Discussion

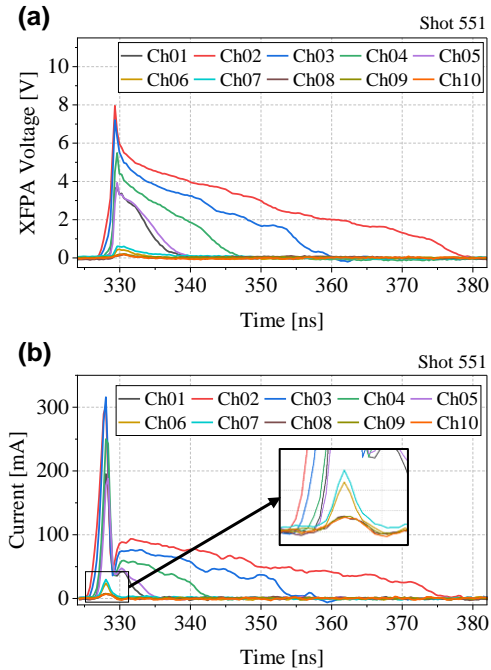

FIG. 10. Application of the reconstruction method to a single thermal x-ray burst (Shot 551). (a) Measured XFPA voltage waveforms. (b) Reconstructed effective source currents. The waveforms indicate a single, short burst lasting less than ~2 ns. Tails are present in the saturated channels (Ch 01–05).

Figure 10(a) shows the measured voltage waveforms from a Cu wire X-pinch (Shot 551). It shows that all channels respond around 328-330 ns. The peak voltage of the low-energy channels (Ch 01–05) exceeds 3V, which is significantly higher than the ~1 V saturation threshold (Sec. III), indicating operation in the nonlinear regime. The effective source currents, reconstructed using Wiener deconvolution (Sec. III), are presented in Figure 10(b). While the already saturated low-energy channels (Ch 01–05) exhibit a tail current, all high-energy channels (Ch 06–10, >2.5 keV) show no significant tail, suggesting they operate within or near the linear regime. This shows that the extended effective source current observed in the low-energy channels after 330 ns reflects saturation of the low-energy channels rather than continuous x-ray emission.

This explains why ratio-based, peak-window analyses can overestimate plasma density when low-energy channels are saturated. For instance, a previous study by Ham et al.[37] analyzed only the peak region of the first x-ray burst (e.g., 329–333 ns for shot 551) to avoid the tail, which led to the estimation of high-density exceeding $10^{23}$ cm$^{-3}$. However, the true incident x-ray flux is proportional to the total collected charge, which includes the contribution from the tail. A forced time-resolved analysis of only the peak region therefore systematically underestimates the source current of the saturated low-energy channels. This suppression of the 1–2 keV flux biases power-ratio-based analyses toward continuum-dominated, high-density solutions. Furthermore, the presence of the tail current could be misinterpreted as a low-energy (~1.5 keV) afterglow,

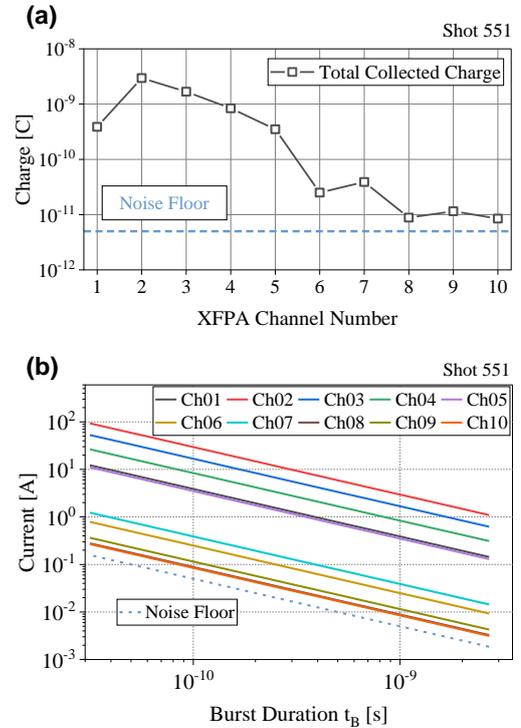

FIG. 11. (a) Total collected charge ($Q_{ch,exp}$) measured for each of the 10 XFPA channels for shot 551. The dashed line indicates the experimentally determined noise floor. (b) Effective channel source currents ($I_{ch,exp} = Q_{ch,exp}/t_B$) versus assumed burst duration ($t_B$) for Shot 551 on a log-log scale.

even when no x-rays are being emitted after the initial single burst.

Figure 11(a) presents the primary data for Shot 551, which is the total collected charge ($Q_{ch,exp}$) measured for each of the 10 XFPA channels. This charge data is then used to derive the effective channel source currents ($I_{ch,exp} = Q_{ch,exp}/t_B$), which are plotted against the assumed burst duration ($t_B$) in Figure 11(b). These lines represent the experimental target data for the parameter estimation. The parallel nature of the lines visually confirms the charge conservation established in Sec. III, indicating that the relative energy deposited in each channel is independent of the assumed $t_B$. The plot also shows the experimentally determined noise floor. The large dynamic range across channels, with some high-energy signals (e.g., Ch 10) only slightly above this floor, emphasizes the importance of the log-transformed least-squares fitting and the use of a noise-floor-based stabilization constant ($\varepsilon$) to ensure that all channels contribute to the fit without undue influence from noise or high-amplitude channels.

The analysis methodology was then applied to the single thermal burst from Shot 551. The assumed burst duration $t_B$ was scanned over a plausible range from 30 ps to 3 ns, and the resulting plasma parameter estimates are presented in Figure 12. The estimated plasma density $n_e$ and electron temperature $T_e$ are relatively insensitive to the choice of $t_B$. This stability indicates that these parameters are constrained by the spectral shape, primarily determined by the ratios of the channel charges. However, the estimated source diameter $d$ exhibits a clear dependence on the assumed $t_B$. This arises because, with $n_e$ and $T_e$ largely constrained by spectral ratios, the diameter $d$ tends to vary to match the absolute luminosity required by the total measured charge and assumed $t_B$. We find that the diameter scales approximately as $d$ is proportional to $t_B^{1/3}$ (more precisely, the exponent is found to be ~1/2.85).

This scaling is different from the dependence of $d$ on $t_B^{1/2}$, which is expected for a surface-emitting optically thick source. The observed scaling suggests that the plasma emission in the 1–10 keV range is predominantly volumetric (closer to optically thin), where total power scales roughly with volume, although the deviation from 1/3 implies that self-absorption effects are not entirely negligible.

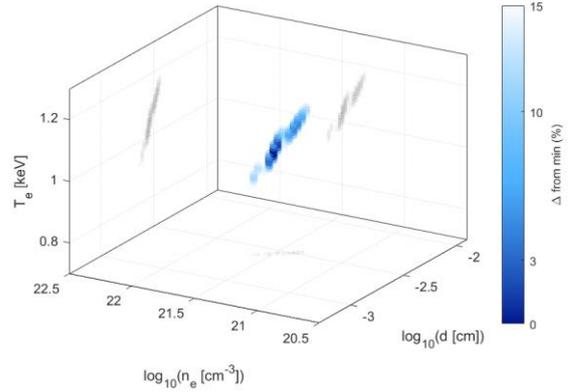

FIG. 13. Visualization of the solution confidence region in the 3D parameter space ($n_e$, $T_e$, $d$) for Shot 551, assuming $t_B = 1$ ns. The color indicates the percentage deviation of the log-transformed least-squares sum from its minimum.

To visualize the solution space around the optimal parameters determined for a given $t_B$, Figure 13 shows the distribution of suboptimal solutions for Shot 551, assuming $t_B = 1$ ns. The color intensity represents the deviation (in percentage) of the log-transformed least-squares sum from its minimum value. The plot illustrates a well-defined valley or cluster of solutions concentrated around the best-fit point. This supports the approach of using the center of mass of these suboptimal points (defined as those within 3% of the minimum LS value in this analysis) as an estimate for the effective plasma parameters ($n_e$, $T_e$, $d$). This 3% is a conservative choice informed by a broader sensitivity analysis performed across multiple shots and the full range of plausible $t_B$ values. This analysis showed that while the primary solution cluster remains dominant and well-isolated, secondary clusters representing other local minima begin to appear at tolerance above ~6%. Thus, the 3% threshold ensures that the analysis is restricted to this single, most probable solution space. Furthermore, the elongated and tilted shape of this solution confidence region visually demonstrates the inherent correlations, or degeneracies, between the parameters in this luminosity-based model. Specifically, it reveals the expected anti-correlations that maintaining a constant total luminosity requires the diameter $d$ to decrease as either the density $n_e$ or the temperature $T_e$ increases.

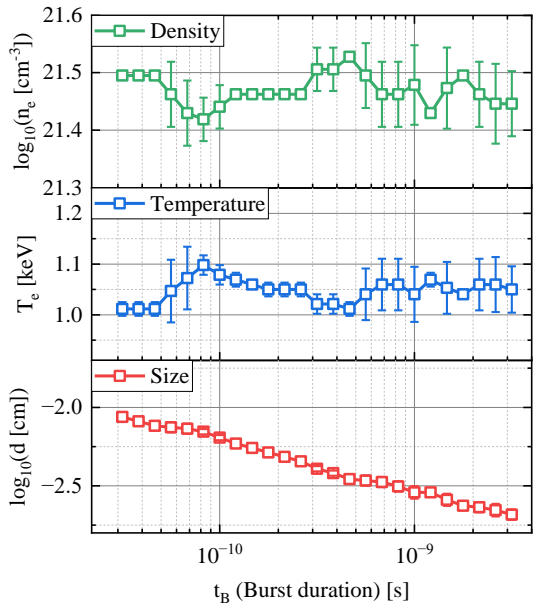

FIG. 12. Estimated plasma parameters as a function of the assumed x-ray burst duration for shot 551. Error bars represent the spread of suboptimal solutions (solution confidence region).

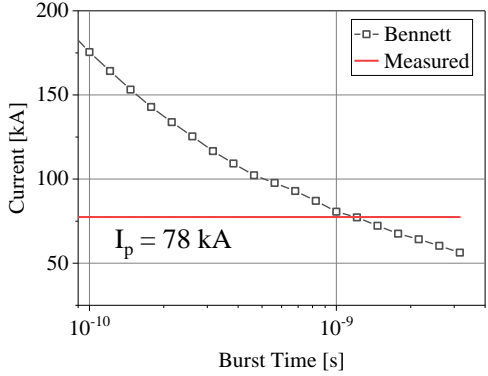

FIG. 14. Determination of the burst duration for Shot 551. The Bennett current $I_{Bennett}(t_B)$ (derived from spectrally inferred parameters) and the experimentally measured pinch current $I_{p,measured}$ (red line) intersects at $t_B \sim 1.2$ ns.

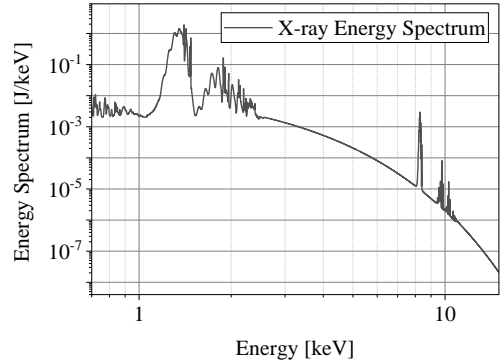

FIG. 15. Reconstructed absolute x-ray energy spectrum for Shot 551, using the determined plasma parameters at $t_B = 1.2$ ns. The total radiated x-ray energy (0.7-28 keV) is estimated to be ~171 mJ.

Figure 14 plots calculated $I_{Bennett}(t_B)$ against the assumed burst duration for Shot 551. Also plotted, as a red line, is the pinch current measured at the time of the SXR burst, $I_{p,measured} \sim 78$ kA. For Shot 551, intersection occurs at $t_B^* \sim 1.2$ ns. This determined $t_B^*$ is then used to specify the final effective plasma parameters for this shot. Therefore, the final parameters characterizing the thermal SXR source in Shot 551 are determined as $\log_{10}(n_e \text{ [cm}^{-3}\text{]}) \sim 21.4$, $T_e \sim 1.1$ keV, $d \sim 29$ μm, and $t_B^* \sim 1.2$ ns.

The determined parameters ($n_e$, $T_e$, $d$, $t_B^*$) allow for the reconstruction of the absolute x-ray energy spectrum emitted during the burst through Equation (2). Figure 15 presents the reconstructed spectrum for the representative Shot 551, plotted in absolute units (J/keV) over a broad energy range pertinent to SXR diagnostics (0.7 keV to >10 keV). This broadband spectrum, derived from our physics-constrained analysis incorporating reasonable luminosity, opacity effects, and Bennett equilibrium, offers a level of detail often missing in previous X-pinch studies that relied on power ratio-based analysis[37] or blackbody assumptions which are inappropriate for optically thin x-ray sources. The spectrum shows that the emission is dominated by Cu L-shell line radiation in the 1–1.5 keV range, accounting for approximately 154 mJ, which constitutes ~90% of the total estimated x-ray energy of ~171 mJ. In contrast, the continuum radiation above 2 keV is significantly weaker, contributing only ~5.5 mJ (~3.2%). This spectral characteristic of strong L-shell emission coupled with a weak continuum is consistent with the inferred plasma density, and suggests that the SXR source is almost optically thin in the range of 1–10 keV.

The spectrum characterized by the weak continuum emission, combined with the effective source diameter of ~29 μm and density of $\log_{10}(n_e \text{ [cm}^{-3}\text{]}) \sim 21.4$ implies that the SXR source produced in the SNU X-pinch under these slow current-rise conditions is a 'bright spot' (BS) rather than an extremely compressed, micron-scale 'hot spot' (HS). This classification aligns with the framework established in X-pinch studies, particularly distinguishing emission regimes based on driver dI/dt characteristics and resulting plasma parameters[39, 49]. Hot spots, typically requiring dI/dt > 1 kA/ns for formation[49], are defined by their extreme parameters: micron or sub-micron size, sub-nanosecond (often ps-scale) duration, near-solid ($n_e \sim 10^{23}$–$10^{24}$ cm$^{-3}$) density, and a spectrum featuring strong continuum radiation alongside lines. In contrast, the parameters derived in this work fit squarely within the bright spot regime.

The effective source size determined by our analysis $d \sim 29$ μm is highly consistent with previous imaging studies of slow-rise X-pinches when observed in softer x-ray bands. For instance, Christou et al.[8] (~ 0.25 kA/ns) measured time-integrated source dimensions of approximately 50×40 μm using aluminum filters sensitive primarily to energies above 1.2 keV. Andola et al.[10] (~ 0.11 kA/ns), using 6 μm Al filter (>1 keV), placed an upper bound of < 25 μm on the source size and noted its suitability for radiography due to the soft-dominant output. It is well-established that the apparent size of the X-pinch emission region depends on the energy bandpass of the diagnostic filter[1, 4, 49]. While a harder filter (> 2–2.5 keV, using Ti filters) tends to isolate a more compact, continuum-dominated core region, often measured to be around ~10–15 μm[7, 9, 13], our analysis is heavily weighted by the dominant 1–1.5 keV Cu L-shell radiation captured by multiple lower-energy XFPA channels. Therefore, the ~30–40 μm effective diameter reflects the characteristic size of this larger, L-shell emitting volume, which is consistent with the bright spot definition[39, 49].

Furthermore, the inferred plasma parameters support this classification. The plasma density $n_e$ is within the range $10^{21}$–$10^{22}$ cm$^{-3}$, typically associated with X-pinch plasmas emitting significantly above 1 keV[13], but remains below the extreme density of hot spots. The electron temperature is also consistent with estimates from other slow-driver experiments, which range from several hundred eV up to ~1.5 keV[8, 13]. The burst duration $t_B$ determined with the Bennett constraint shows clear agreement with direct measurements on similar portable capacitor-driven systems, such as the 1.1±0.3 ns FWHM reported by Strucka et al.[7] (~0.45 kA/ns) and the 0.4–0.7 ns FWHM observed by Aranchuk et al[13]. The slightly smaller source size reported by Strucka et al.[7] (~10 μm) can be attributed to their potentially harder effective bandpass filtering and moderately higher dI/dt, both factors contributing to resolving a more compact core.

| Shot Number | Current Rise Rate [kA/ns] | Wire Diameter [μm] | Pinch Current [kA] | $\log_{10}(n_e \text{ [cm}^{-3}\text{]})$ | $T_e$ [keV] | d [μm] | $t_B$ [ns] | Total X-ray Energy [mJ] | X-ray Energy (> 2 keV) [mJ] |
|---|---|---|---|---|---|---|---|---|---|
| 551 | 0.24 | 30 | 78 | 21.4 | 1.1 | 29 | 1.2 | 171 | 5.5 (3.2%) |
| 676 | 0.32 | 25 | 87 | 21.2 | 1.1 | 42 | 0.8 | 168 | 3.9 (2.3%) |
| 588 | 0.27 | 15 | 57 | 21.1 | 1.1 | 31 | 1.2 | 75 | 1.6 (2.1%) |
| 566 | 0.19 | 15 | 51 | 20.9 | 1.3 | 37 | 1.5 | 48 | 1.2 (2.5%) |
| 574 | 0.12 | 15 | 43 | 20.8 | 1.1 | 36 | 1.0 | 29 | 0.3 (1.0%) |
| 573 | 0.12 | 15 | 43 | 20.9 | 1.5 | 31 | 1.0 | 14 | 0.4 (2.8%) |

TABLE. 1. Summary of experimental conditions and determined plasma parameters for representative single thermal bursts for Cu-wire X-pinch experiments. Parameters include current rise rate (dI/dt), wire diameter, pinch current ($I_p$), derived electron density ($n_e$), electron temperature ($T_e$), source diameter (d), burst duration ($t_B$), total radiated SXR energy (0.7–28 keV), and energy radiated above 2 keV.

The total x-ray energy of ~171 mJ is also consistent with yields reported from other portable, capacitor-driven X-pinches when considering the typical variations due to driver parameters, load configurations, and, crucially, the diagnostic energy bandpass. For example, Strucka et al.[7] measured ~57 mJ in the 1–10 keV band and Aranchuk et al.[13], operating at 200 kA, reported $O(10^2 \text{ mJ})$ yields specifically in the SXR range (>800 eV) for both Cu and Mo wires. Studies at even lower dI/dt (~0.11 kA/ns) by Andola et al.[10] also showed soft-band yields in the tens of mJ range (e.g., 25–40 mJ for >1–1.5 keV), while explicitly demonstrating the strong dependence of measured yield on the filter cutoff energy. While direct comparison requires careful consideration of these differing conditions, the energy scale observed in our experiments fits within the envelope typically associated with soft x-ray yields generated on slow-rise drivers.

To assess the generality of these findings beyond the representative Shot 551, the same analysis procedure was applied to several other single thermal burst events recorded under varying experimental conditions. The key results for six representative shots, including different current rise rates dI/dt, wire diameters, and measured pinch currents, are summarized in Table 1. Across all analyzed shots, the derived parameters consistently fall within the 'bright spot' regime established above, and the burst durations remain on the nanosecond scale ($t_B$ ~ 0.8–1.5 ns). This consistency supports our conclusion that bright spot formation is the characteristic SXR generation mechanism for the SNU X-pinch, which is operating in this slow current-rise regime.

Table 1 reveals several trends related to the driving conditions. The most evident trend is the strong positive correlation between the measured pinch current ($I_p$) and the total radiated soft x-ray energy. A higher pinch current is consistently observed to produce a higher X-ray yield. In contrast, the electron temperature ($T_e$) remains relatively uniform across all shots near ~1.1 keV, and the burst duration ($t_B$) also shows consistency at approximately ~1 ns.

A more detailed examination of the dI/dt and wire diameter effects is notable, particularly the comparison between Shot 551 (30 μm) and Shot 588 (15 μm), which had similar dI/dt rates (~0.24–0.27 kA/ns). Using a thicker wire, Shot 551 achieved a 1.37 times higher pinch current, leading to an increase of 2.3 times of the total x-ray energy. The x-ray energy above 2 keV increased even more significantly, by a factor of 3.4, due to the two times higher plasma density achieved in Shot 551 ($\log_{10}(n_e \text{ [cm}^{-3}\text{]})$ ~ 21.4) compared to Shot 588 ($\log_{10}(n_e \text{ [cm}^{-3}\text{]})$ ~ 21.1), enhancing the relative contribution of continuum radiation. This relationship between plasma density and the fraction of >2 keV x-ray energy appears consistent across the dataset. For instance, Shot 551, with the highest density ($\log_{10}(n_e \text{ [cm}^{-3}\text{]})$ ~ 21.4), shows a >2 keV energy fraction of 3.2%, whereas Shot 574, with one of the lowest densities ($\log_{10}(n_e \text{ [cm}^{-3}\text{]})$ ~ 20.8), has a fraction of only 1.0%. Importantly, even under the highest current and dI/dt conditions achieved in this study (e.g., Shot 676), the fraction of energy radiated above 2 keV remains low (typically 1–3.2%). This observation reinforces that while the overall brightness increases, the fundamental nature of the source as an L-shell dominant 'bright spot' persists within this operational regime of the SNU X-pinch.

## VI. Conclusion

This work addresses a diagnostic problem for pulsed power experiments, specifically the interpretation of silicon PIN photodiode signals. We have shown that while the temporal profile of a saturated signal is severely distorted, the total collected charge remains a conserved quantity proportional to the incident pulse energy. This finding enables spectroscopic analysis that relies on this charge conservation. By applying this finding within a framework that combines a spherical radiation model and the Bennett relation, we perform parameter estimation and spectroscopy for the thermal x-ray burst source. Our analysis indicates that the source in the low dI/dt regime is a 'bright spot' ($n_e$ ~ $10^{21}$ cm$^{-3}$, d ~ 30–40 μm, $T_e$ ~ 1 keV, $t_B$ ~ 1 ns), a finding that is physically distinct from the formation of an extremely compressed 'hot spot'. This result addresses previous ambiguities in the characterization of low dI/dt experiments, which were often complicated by photodiode nonlinear effects such as tail current and peak suppression. Furthermore, this methodology, which relies on total charge rather than distorted temporal information, offers a practical framework for quantitative spectroscopy in other current-driven pulsed-power systems, such as Z-pinches or plasma focus devices, where diagnostic saturation under intense radiation remains a common experimental limitation.

**ACKNOWLEDGMENTS**

This work was supported by the Korea Research Institute for Defense Technology Planning and Advancement (KRIT) grant funded by the Korea government (DAPA) (No. KRIT-CT-25-002, Dense Z-pinch (DZP) Plasma Research Laboratory) and by the Defense Research Laboratory Program of the Defense Acquisition Program Administration and Agency for Defense Development of the Republic of Korea, and the National Research Foundation of Korea (NRF) grant funded by the Korea government (MSIT) (Grant No. RS-2023-00208337).

**AUTHOR DECLARATIONS**

**Conflict of Interest**

The authors have no conflicts to disclose.